\documentclass[aps,prb,onecolumn,floatfix]{revtex4}

\usepackage{epsf}
\usepackage{subfigure}
\usepackage{amsmath}
\usepackage{amssymb}
\usepackage{bm}
\usepackage{graphicx}
\usepackage{epstopdf}
\newcommand{\be}{\begin{equation}}
\newcommand{\ee}{\end{equation}}
\newcommand{\bea}{\begin{eqnarray}}
\newcommand{\eea}{\end{eqnarray}}

\newcommand{\parent}[1]{\left( #1 \right)}

\begin{document}

\title{\bf Solving bifurcation diagrams using fixed points as control parameters}

\author{David Andrieux}

\begin{abstract}
We propose to determine the bifurcation diagrams of fixed points using their coordinates as control parameters.
This method can lead to exact solutions to otherwise intractable bifurcation problems.
\end{abstract}

\maketitle

\vskip 0,25 cm

\section*{EXECUTIVE SUMMARY}

We show that bifurcations diagrams can be obtained by using the dynamical variables as control parameters. 
Instead of looking at the evolution of a fixed point as a function of the system parameters, we deduce the system parameters as a function of the fixed point coordinates.

This approach has several advantages. 
In many cases analytical solutions can be obtained as it {\it bypasses the need to solve a nonlinear system}. 
Furthermore, by construction, it provides a {\it global analysis}. The whole bifurcation diagram is obtained at one go.

The key insight behind this result is that the system parameters are usually coupled linearly to the dynamics. 
Therefore, given a fixed point, the bifurcation problem is reduced to solving a set of linear equations for the system parameters.
The whole bifurcation diagram is then constructed by varying the fixed point coordinates.


We introduce the technique through examples. A similar reasoning can then be applied to other dynamical systems of interest.

\vskip 0,75 cm

\section{Example 1. FitzHugh-Nagumo model}

Our first example is the FitzHugh-Nagumo model
\begin{subequations}
\label{FH}
\bea
\frac{{\rm d}v}{{\rm d}t} &=& v - v^3 -w + I \\
\tau \frac{{\rm d}w}{{\rm d}t} &=& v - aw \, .
\eea
\end{subequations}
The FitzHugh-Nagumo describes a prototype of an excitable system (e.g., a neuron).

We use this simple example to illustrate how to simplify the bifurcation analysis by avoiding solving a cubic nonlinearity for each values of the parameters $I$ and $a$. 

The key observation is that $I$ and $a$ are coupled linearly to the dynamical variables $v$ and $w$. 
Therefore, a fixed point $(\bar{v},\bar{w})$ will impose a set of linear constraints for the parameters $(I,a)$. 
Solving this linear system leads to
\begin{subequations}
\bea
I (\bar{v},\bar{w}) &=& \bar{w}-\bar{v}+\bar{v}^3 \\
a (\bar{v},\bar{w}) &=& \bar{v}/\bar{w} \, . 
\eea
\end{subequations}

These equations give the combination $(I, a)$ for which $(\bar{v},\bar{w})$ is a fixed point of (\ref{FH}).
We can now obtain the whole bifurcation diagram analytically by varying the fixed point $(\bar{v},\bar{w})$ and calculating its stability.
Notably, this is achieved by solving a linear system rather a cubic nonlinearity as in the standard approach.

\section{Example 2. Bistability in population networks}

We now consider a system of two variables $x_{1,2} \in [0,1]$ interacting through a positive, increasing function $F$:
\begin{subequations}
\label{network}
\bea
\frac{{\rm d}x_1}{{\rm d}t} &=& - \frac{x_1}{\tau} + \gamma (1-x_1) F\parent{W_{11} x_1 + W_{12} x_2 } \\
\frac{{\rm d}x_2}{{\rm d}t} &=& - \frac{x_2}{\tau} + \gamma (1-x_2) F\parent{W_{21} x_1 + W_{22} x_2 } \, ,
\eea
\end{subequations}
where the coupling coefficients satisfy 
\bea
W_{11}=W_{22} \quad {\rm and} \quad W_{12}=W_{21} \, .
\eea
Such dynamical systems appear, for example, in mean-field descriptions of neural networks (see, e.g., Ref. \cite{WW06}).

This system admits at least one symmetric fixed point $(x_0, x_0)$. 
Depending on the coupling strengths, the network can also admit asymmetric states. 
However, none of the fixed points (symmetric or asymmetric) can be obtained analytically.

We use our method to study the bistability properties of the network (\ref{network}) in terms of the coefficients $W_{ij}$. 
The interaction between the dynamical variables $x_i$ and the $W$s is linear.
Therefore we can determine the $W$s compatible with a fixed point $(\bar{x}_1, \bar{x}_2)$.
Noting that $F$ is invertible since it is increasing, we have that
\bea
\left(\begin{array}{cc}
   \bar{x}_1 & \bar{x}_2     \\
   \bar{x}_2 & \bar{x}_1   \\
\end{array} \right) 
\left(\begin{array}{c}
   W_{11} \\
   W_{12} \\
\end{array} \right) 
= 
\left(\begin{array}{c}
K_1  \\ 
K_2  \\ 
\end{array} \right) \, ,  \nonumber
\eea
where 
\bea
K_i =  F^{-1} \parent{ \frac{\bar{x}_i}{\gamma \tau (1-\bar{x}_i)} }  \, . \nonumber
\eea
The solution of this system reads
\begin{subequations}
\label{W}
\bea
W_{11} (\bar{x}_1,\bar{x}_2) = W_{22} (\bar{x}_1,\bar{x}_2) &=& \frac{\bar{x}_1K_1 - \bar{x}_2K_2}{(\bar{x}_1+\bar{x}_2)(\bar{x}_1-\bar{x}_2)} \label{W11}\\
W_{12} (\bar{x}_1,\bar{x}_2) = W_{21} (\bar{x}_1,\bar{x}_2) &=& \frac{\bar{x}_1K_2 - \bar{x}_2K_1}{(\bar{x}_1+\bar{x}_2)(\bar{x}_1-\bar{x}_2)} \label{W12}
\eea
\end{subequations}
when $\bar{x}_1 \neq \bar{x}_2$.

\begin{figure}[h]
\centerline{\includegraphics[width=10cm]{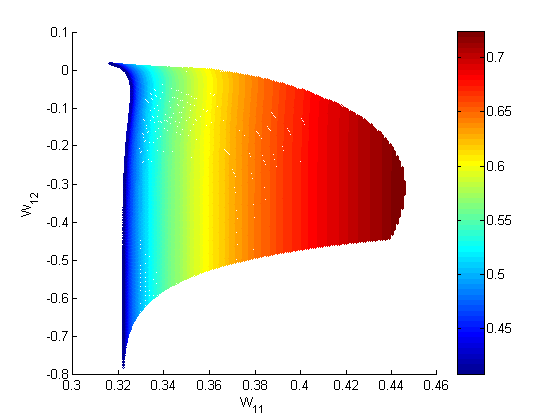}} 
\caption{{\bf  Bistability region of the population network (\ref{network}).} 
The colored region depicts the couplings for which symmetric and asymmetric stable fixed points coexist.  
The color code indicates the value $\max (\bar{x}_1,\bar{x}_2)$ of the asymmetric state. 
The function $F$ as well as the values of $\gamma$ and $\tau$ used in this figure can be found in Ref. \cite{WW06}. 
The coupling coefficients are determined by Eqs. (\ref{W}).}
\label{fig1}
\end{figure}

Equations (\ref{W}) express the coupling parameters in terms of the fixed point coordinates $(\bar{x}_1,\bar{x}_2)$. 
We can now obtain the whole bistability region analytically by varying these coordinates.
This bistability region is depicted in Fig~\ref{fig1}.

Note that, in this example, the standard problem of finding the fixed points as a function of the coupling is not analytically tractable. 
Nonetheless, our approach allows us to obtain the complex bifurcation diagram seen in Fig \ref{fig1} exactly. 
Furthermore, it gives the global solution, i.e. the whole diagram, at one go. 
In particular, there is no need to follow different fixed points or branches to reconstruct the diagram.

\newpage

\vskip 0,6 cm

{\bf Acknowledgments.} Part of this work was done at Yale University. 

\vskip 0,5 cm

{\bf Disclaimer.} This paper is not intended for journal publication.


\end{document}